\documentclass[twoside]{ilcws08}
\usepackage[latin1]{inputenc}
\usepackage[dvips]{graphicx,epsfig,color}
\usepackage{wrapfig,rotating}
\usepackage{amssymb,amsmath,array}
\usepackage{hyperref}
\newcommand{\ep}{\varepsilon}
\newcommand{\Li}[2]{{\mbox{Li}}_{#1}\left(#2\right)}
\begin{document}
\title{
Differential Reduction Algorithms for
Hypergeometric Functions Applied to Feynman Diagram Calculation} 
\author{V.V. Bytev$^1$\thanks{Research supported by MK 1607.2008.2.}, 
M. Kalmykov$^{1,2}$
\thanks{Research supported by BMBF 05 HT6GUA.}, 
B.A. Kniehl$^2$, B.F.L. Ward$^3$, and 
S.A. Yost$^4$\thanks{Speaker. See Ref.\ \cite{url} for the presentation slides.}
\vspace{.3cm}\\
1- Joint Institute for Nuclear Research, \\
$141980$ Dubna (Moscow Region), Russia \\
\vspace{.1cm}\\
2- II. Institut f\"ur Theoretische Physik, Universit\"at Hamburg, \\
Luruper Chaussee 149, 22761 Hamburg, Germany \\
\vspace{.1cm}\\
3- Department of Physics, Baylor University, \\
 One Bear Place, Waco, TX 76798, USA \\
\vspace{.1cm}\\
4- Department of Physics, The Citadel,\\
171 Moultrie St., Charleston, SC 29409, USA
}
\maketitle

\begin{abstract}
We describe the application of differential reduction algorithms for 
Feynman Diagram calculation.  We illustrate the procedure in the context of 
the generalized hypergeometric functions ${}_{p+1}F_{p}$, and give an
example for a type of $q$-loop bubble diagram. 
\end{abstract}

\section{Introduction}
A Feynman diagram can be understood mathematically as a linear combination of 
Horn-type hypergeometric functions of several variables 
(see Ref.\ \cite{acat08}):
\begin{eqnarray}
\sum_j B_j x_1^{\alpha_1} \cdots x_r^{\alpha_r} \Phi(\vec{\gamma},\vec{\sigma},\vec{x}) \;, 
\label{fd}
\end{eqnarray}
where a Horn-type hypergeometric function has the structure
\begin{eqnarray}
\Phi(\vec{\gamma};\vec{\sigma};\vec{x}) 
= 
\sum_{m_1,m_2,\cdots, m_r=0}^\infty 
\Biggl( 
\frac{
\prod_{j=1}^K
\Gamma\left( \sum_{a=1}^r \mu_{ja}m_a+\gamma_j \right)
}
{
\prod_{k=1}^L
\Gamma\left( \sum_{b=1}^r \nu_{kb}m_b+\sigma_k \right)
}
\Biggr) 
x_1^{m_1} \cdots x_r^{m_r} \;,
\label{Phi}
\end{eqnarray}
with the arguments $x_j$ being, in general, rational functions
(typically, simple ratios) of kinematic invariants of the original Feynman
diagram, and the parameters $\{ \gamma_j \}$ and $\{ \sigma_k \}$ 
being linear combinations of the exponents of
propagators and the dimension of space-time.\footnote{The
presence of a nontrivial numerator in the Feynman diagram does not affect
this conclusion \cite{numerator}.} The $\gamma_j$ and $\sigma_k$ are 
called upper and lower parameters, respectively.

These statements follow from the multiple Mellin-Barnes representation for
a dimension\-al\-ly regularized Feynman diagram (see Ref.\ \cite{Smirnov}),
 and the assumption that there is a region of variables where every
term in the linear combination (\ref{fd}) is convergent.
The Horn-type structure permits the hypergeometric functions appearing in
(\ref{fd}) to be reduced to a set of basis functions with parameters differing
from the original ones by integer shifts:
\begin{eqnarray}
P_0(\vec{x})
\Phi(\vec{\gamma}+\vec{l};\vec{\sigma}+\vec{s};\vec{x})
= 
\sum_{m_1, \cdots, m_r=0}^{\sum_j|l_j|+\sum_k|s_k|} 
P_{r_1, \cdots, r_p} (\vec{x}) D_1^{m_1} \cdots D_r^{m_r}
H(\vec{\gamma};\vec{\sigma};\vec{x}) \;,
\label{dr}
\end{eqnarray}
where $D_j^{r} = \left( \frac{\partial}{\partial x_j} \right)^r$ denotes a
partial derivative and 
$P_{r_1, \cdots, r_p}(\vec{x})$ are rational functions \cite{Takayama}.
These shifts may be implemented by constructing a set of four differential 
operators 
 $U_{\gamma_c}^\pm$, $L_{\sigma_c}^\pm$  which respectively change
$\gamma_c$, $\sigma_c$ by $\pm 1$:  $\gamma_c\rightarrow \gamma_c \pm 1$ or 
$\sigma_c \rightarrow \sigma_c \pm 1$.  These basic operators are called the
step-up and step-down operators for the upper and lower parameters.
A procedure of applying step-up and step-down operators to reduce the original
hypergeometric function to a basis set is called a {\bf differential reduction}.
In the case when some of the variables are equal to one another,
$x_i=x_j,i\neq j$, or belong to the surface of singularities
$Q = \{\vec{x}| P_0(\vec{x})=0\}$,
it is necessary to define a limiting procedure for (\ref{dr}).

We will illustrate our approach by considering the reduction of a 
particular generalized hypergeometric function in section \ref{generalized}, 
and by applying it to a particular class of Feynman diagrams in section 
\ref{example}. Section \ref{hyperlogarithms} describes the 
reduction at a singular surface using the $\ep$ expansion and hyperlogarithms.

\section{Generalized hypergeometric function of one variable}
\label{generalized} 

In this section, we will show how differential reduction may be applied to a 
generalized hypergeometric function of one variable.
Let us recall that the generalized hypergeometric function $_pF_{p-1}(a;b;z)$ 
may be defined in a neighborhood of $z=0$ by the series 
\begin{equation}
F(\vec{a};\vec{b};z) 
\equiv
{}_{p}F_{p-1} \left( \begin{array}{c|}
\vec{a} \\
\vec{b}
\end{array}~ z \right)
= \sum_{k=0}^\infty \frac{z^k}{k!} 
\frac{\prod_{i=1}^p (a_i)_k}{\prod_{j=1}^{p-1} (b_j)_k} \;,
\label{hypergeometric}
\end{equation}
where 
$(a)_k = \Gamma(a+k)/\Gamma(a)$ is called a
Pochhammer symbol.  The lists $\vec{a}=(a_1,\cdots, a_p)$
and $\vec{b}=(b_1,\cdots, b_q)$ are called upper and lower parameters of the 
hypergeometric function, respectively. 
The hypergeometric function ${}_pF_{p-1}$ satisfies a differential equation
\begin{equation}
L(\vec{a},\vec{b}) \left( {}_pF_{p-1}(\vec{a}; \vec{b}; z) \right) 
= 
\left[ 
z \prod_{i=1}^p (\theta \!+\! a_i) \!-\! \theta 
\prod_{i=1}^{p-1} (\theta \!+\! b_i-1)
\right] 
{}_pF_{p-1}(\vec{a}; \vec{b}; z) = 0 \; ,
\label{diff}
\end{equation}
where $L(\vec{a},\vec{b})$ is a differential operator and 
$\theta = z \frac{d}{d z}$ \;. 

Constructing a reduction scheme requires a set of step-up and step-down
differential operators for both the upper and lower parameters. 
In this case, the universal step-up (step-down) operators for the 
upper (lower) parameters have a very simple form:\footnote{See 
Ref.\ \cite{Takayama}, or Eq.\ (2.1), (2.2) in Ref.\ \cite{acat08}, or 
Ref.\ \cite{Rainville} for details.}
$$
U_{a_i}^+  =    \frac{\theta+a_i}{a_i} \; , \quad 
L_{b_j}^-  =    \frac{\theta+b_j-1}{b_j-1} \;, 
$$
and the inverse operators 
$U_{\vec{a}}^-$, and $L_{\vec{b}}^+$ can be constructed in accordance with
Takayama's algorithm. \cite{Takayama} (See also Ref.\ \cite{acat08}.)

The differential reduction algorithm takes the form of a product of several 
differential step-up/step-down operators
$U^{\pm}_{\vec{a}},L^{\pm}_{\vec{b}}$: 
\begin{equation}
{}_{p+1}F_p(\vec{a}+\vec{m}; \vec{b}+\vec{n}; z)  = 
\left( U_{\vec{a}}^{\pm} \right)^{\sum_i m_i} 
\left( L_{\vec{b}}^{\pm} \right)^{\sum_j n_j} 
{}_{p+1}F_p(\vec{a}; \vec{b}; z) \;,
\label{general}
\end{equation}
so that the maximal power of $\theta$ in this expression is
$r \equiv \sum_i m_i+\sum_j n_j$.  Since the hypergeometric function  
${}_{p+1}F_{p}(\vec{a};\vec{b}; z)$ satisfies a differential 
equation (\ref{diff}) of order $p$,
it is possible to express all terms containing powers of $\theta^k$ 
with $k \geq p$
in terms of 
$\theta^j {}_{p+1}F_p(\vec{a};\vec{b};z)$  with $j \leq p$, 
multiplied by coefficients that are rational functions of the 
parameters and the argument $z$.  In this way, any function  
${}_{p+1}F_{p}(\vec{a}+\vec{m};\vec{b}+\vec{k}; z)$ may be expressed in terms 
of a basic function ${}_{p+1}F_{p}(\vec{a};\vec{b}; z)$ and its first 
$p$ derivatives: 
\begin{eqnarray}
\label{decomposition}
&& \hspace{-5mm}
S(a_i,b_j,z)
{}_{p+1}F_{p}(\vec{a}+\vec{m};\vec{b}+\vec{k}; z)
 = 
\\ && \hspace{-5mm}
\Biggl \{
  R_1(a_i,b_j,z) \theta^p
+ R_2(a_i,b_j,z) \theta^{p-1}
+ \cdots 
+ R_p(a_i,b_j,z) \theta
+ R_{p+1}(a_i,b_j,z) 
\Biggr\}
{}_{p+1}F_{p}(\vec{a};\vec{b}; z) \;,
\nonumber 
\end{eqnarray}
where 
$\vec{m},\vec{k}$ are lists of integers  and 
$S$ and $T_i$ are polynomials in the parameters $\{a_i\},\{b_j\}$ and $z$.
For some special sets of parameters, the result of the reduction 
(\ref{decomposition}) takes a simple form. 
In particular, when one of the upper parameters is an integer, 
\begin{eqnarray}
&& \hspace{-5mm}
\tilde{S}(a_i,b_j,z)
{}_{p+1}F_{p}(\vec{j},\vec{a}\!+\!\vec{m};\vec{b}\!+\!\vec{k}; z)
 = 
\nonumber \\ && \hspace{-5mm}
  \tilde{R}_1(a_i,b_j,z) 
+ 
\Biggl \{
  \tilde{R}_2(a_i,b_j,z) \theta^{p-1}
+ \cdots 
+ \tilde{R}_p(a_i,b_j,z) \theta
+ \tilde{R}_{p+1}(a_i,b_j,z) 
\Biggr\}
{}_{p+1}F_{p}(\vec{a};\vec{b}; z) \;.
\nonumber 
\end{eqnarray}
For further details see Ref.\ \cite{BKK}.\\
\begin{wrapfigure}{r}{5.0cm}
\vspace{-1cm}
\centerline{\includegraphics[width=4.0cm]{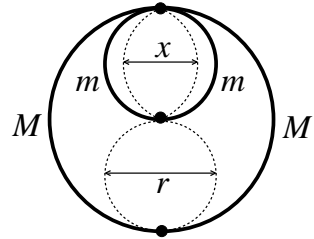}}
\vspace{-0.2cm}
\caption{$q$-loop bubble $B^q_{112200}$}\label{Example}
\end{wrapfigure} 
\section{Example of reduction}
\label{example}
As an example of how the differential reduction 
applies\footnote{See also Ref.\ \cite{quark08} for more examples.} 
to a particular type of 
Feynman diagram, let us consider the $q$-loop bubble diagram $B^q_{112200}$ 
in Fig.~\ref{Example},
with four massive lines (with masses $m$, $M$ as indicated) and  
two sets of massless subloops with $r$ and $x$ lines,
respectively.  It is defined as 
\begin{eqnarray}
&& \hspace{-15mm}
B^q_{112200} (m^2, M^2, \alpha_1, \alpha_2, \beta_1, \beta_2, \sigma_1, 
\cdots, \sigma_x, \rho_1, \cdots, \rho_r) \nonumber \\ && 
= 
\int \frac{d^n (k_1 \cdots k_q)}
{
[k_1^2]^{\rho_1} \cdots [k_{r-1}^2]^{\rho_{r-1}} 
[(k_1 \!+\! \cdots \!+\! k_{r-1}  \!+\! P)^2]^{\rho_r}  
}
\nonumber \\ && \hspace{15mm}
\times
\frac{1}{
[k_{r+x}^2-m^2]^{\alpha_1} 
[(k_r \!+\! \cdots \!+\! k_{r+x}  \!+\! P)^2-m^2]^{\alpha_2} 
} 
\\ && \hspace{15mm}
\times
\frac{1}{
[k_r^2]^{\sigma_1} \cdots [k_{r+x-1}^2]^{\sigma_x} 
[k_{q-1}^2-M^2]^{\beta_1} 
[k_{q}^2-M^2]^{\beta_2} 
}
\;, 
\nonumber 
\end{eqnarray}
where 
$
P=k_{q-1}+k_q \;
$
and 
$
x+r=q-2\;, q \geq 3
$
The Mellin-Barnes representation for this diagram is\footnote{We note that a 
smooth limit exists for $\sigma \to 0$ and $\rho \to 0$.}
\begin{eqnarray}
&& \hspace{-15mm}
B^q_{112200} (m^2, M^2, \alpha_1, \alpha_2, \beta_1, \beta_2, \sigma_1, 
\cdots, \sigma_x, \rho_1, \cdots, \rho_r)
\nonumber \\ && 
= 
\frac{
\left[ i^{1-n} \pi^\frac{n}{2} \right]^q
 (-m^2)^{\tfrac{n}{2}\!-\!\alpha_1\!-\!\alpha_2}
 (-M^2)^{\tfrac{n}{2}(q\!-\!1)\!-\!\rho\!-\!\sigma\!-\!\beta_1\!-\!\beta_2}
}
{\Gamma(\alpha_1) \Gamma(\alpha_2) \Gamma(\beta_1) \Gamma(\beta_2) \Gamma\left( \frac{n}{2} \right)}
\nonumber \\ && \hspace{15mm}
\times
\left\{ \prod_{i=1}^{x} \frac{\Gamma(\tfrac{n}{2}-\sigma_i)}{\Gamma(\sigma_i)} \right\}
\left\{ \prod_{j=1}^{r} \frac{\Gamma(\tfrac{n}{2}-\rho_j)}{\Gamma(\rho_j)} \right\}
\frac{ \Gamma\left(\rho-\tfrac{n}{2}(r-1)\right)}{ \Gamma\left(\tfrac{n}{2}r-\rho \right)}
\nonumber \\ && \hspace{15mm}
\times
\int ds 
\left( \frac{M^2}{m^2}\right)^s 
      \Gamma\left(\sigma\!-\!\tfrac{n}{2}x \!-\!s\right) 
      \Gamma\left(\beta_1\!+\!\beta_2\!+\!\rho\!+\!\sigma\!-
	\!\tfrac{n}{2}(q\!-\!1)\!-\!s\right) 
\nonumber \\ && \hspace{15mm}
\times
\frac{\Gamma(\alpha_1\!+\!s) \Gamma(\alpha_2\!+\!s) 
      \Gamma(\alpha_1\!+\!\alpha_2\!-\!\tfrac{n}{2}\!+\!s) 
      \Gamma(\tfrac{n}{2}\!+\!s) 
      \Gamma(\tfrac{n}{2}(q\!-\!2) \!-\! \rho \!-\! \sigma \!+\!s) 
      }
     {
     \Gamma(\alpha_1\!+\!\alpha_2\!+\!2s) 
     \Gamma(\tfrac{n}{2}(x\!+\!1)\!-\!\sigma\!+\!s) 
     }
\nonumber \\ && \hspace{15mm}
\times
\frac{
      \Gamma(\beta_1\!+\!\rho\!+\!\sigma\!-\!\tfrac{n}{2}(q\!-\!2)\!-\!s) 
      \Gamma(\beta_2\!+\!\rho\!+\!\sigma\!-\!\tfrac{n}{2}(q\!-\!2)\!-\!s) 
}
{
     \Gamma(\beta_1\!+\!\beta_2\!+\!2 \sigma \!+\! 2 \rho \!-
	\! n(q\!-\!2) \!-\! 2 s)
}\;, 
\label{B112200MB}
\end{eqnarray}
where we have introduced the notations
$
\rho = \sum_{a=1}^r \rho_a \;, \
\sigma = \sum_{a=1}^x \sigma_a \;. 
$

Closing the contour of integration on the left (see Ref.\ \cite{BD} for 
details), the result may be expressed as a sum of four hypergeometric functions 
of type $_7F_6$.  The general result is too long to reproduce here. 
In the case that the exponents of the propagators are all integers, the 
hypergeometric functions appearing in the result are reducible to the 
following hypergeometric functions and derivatives thereof:\footnote{Two of 
the original hyergeometric functions have a similar 
parameter structure and the same basis functions.}
\begin{eqnarray}
&& 
\{ 1, \theta, \theta^2, \theta^3 \} \times
~_{4}F_3\left(\begin{array}{c|}
I_1 \!-\! \frac{n}{2}(x\!-\!1), 
I_2 \!-\! \frac{n}{2}x, 
I_3 \!-\! \frac{n}{2}(x\!+\!1), 
\frac{1}{2}\!+\!I_4 \!+\! \frac{n}{2}(q\!-\!x\!-\!2) \\ 
I_5\!+\! \frac{n}{2},
I_6 \!+\! \frac{n}{2}(q\!-\!x\!-\!1), 
\frac{1}{2} \!+\! I_7 \!-\! \frac{n}{2}x 
\end{array} ~z \right) 
\nonumber \\ && 
\{ 1, \theta, \theta^2, \theta^3 \} \times
~_{4}F_3\left(\begin{array}{c|}
\frac{1}{2}\!+\!I_1 \!-\! \frac{n}{2},
I_2 \!-\! \frac{n}{2}(q\!-\!2), 
I_3 \!-\! \frac{n}{2}(q\!-\!1), 
I_4 \!-\! \frac{n}{2}q \\ 
I_5 \!-\! \frac{n}{2}(q\!-\!x\!-\!1),
I_6 \!-\! \frac{n}{2}(q\!-\!x\!-\!2), 
\frac{1}{2} \!+\! I_7 \!-\! \frac{n}{2}(q\!-\!1) 
\end{array} ~z \right) 
\label{part2}
\nonumber \\ && 
\{ 1, \theta, \theta^2, \theta^3 \} \times
~_{5}F_4\left(\begin{array}{c|}
1, 
\frac{1}{2}\!+\!I_1,
I_3 \!-\! \frac{n}{2}(q\!-\!1), 
I_2 \!-\! \frac{n}{2}(q\!-\!2), 
I_4 \!-\! \frac{n}{2}(q\!-\!3) \\ 
I_5 \!+\! \frac{n}{2}, 
\frac{1}{2}\!+\!I_6 \!-\! \frac{n}{2}(q\!-\!2), 
I_7\!-\! \frac{n}{2}(q\!-\!x\!-\!2),
I_8 \!-\! \frac{n}{2}(q\!-\!x\!-\!3) 
\end{array} ~z \right) \;,
\label{basis}
\nonumber \\
\end{eqnarray}
where $I_k$ are arbitrary integers. 
In the last expression, some polynomials are also generated. 

Further simplification is possible when $q,x$ take particular values. 
For example, for $x=0$ and $x=1$, the first hypergeometric function is
reducible to 
\begin{eqnarray}
& & \{ 1, \theta \} \times
~_{3}F_2\left(\begin{array}{c|}
1,
I_2 \!-\! \frac{n}{2}, 
\frac{1}{2}\!+\!I_3 \!+\! \frac{n}{2}(q\!-\!2) \\ 
  I_4 \!+\! \frac{n}{2}(q\!-\!1), 
\frac{1}{2} \!+\! I_5
\end{array} ~z \right) \quad (\hbox{for }x=0)\;,
\end{eqnarray}
or  to 
\begin{eqnarray}
& & \{ 1, \theta, \theta^2 \} \times
~_{4}F_3\left(\begin{array}{c|}
1, 
I_1 \!-\! \frac{n}{2}, 
I_2 \!-\! n, 
\frac{1}{2}\!+\!I_3 \!+\! \tfrac{n}{2}(q\!-\!3) \\ 
  I_4\!+\! \frac{n}{2},
I_5 \!+\! \frac{n}{2}(q\!-\!2), 
\frac{1}{2} \!+\! I_6 \!-\! \frac{n}{2}
\end{array} ~z \right) \quad (\hbox{for }x=1)\;.
\end{eqnarray}

\section{Reduction at a singular surface via the $\ep$ expansion and 
hyperlogarithms}
\label{hyperlogarithms}

In physical applications, the case of of equal masses $m^2=M^2$  (a 
``single-scale'' diagram) is of special interest.\footnote{
See Ref.\ \cite{david} for the three-loop case $(q=3,x=0,r=0)$ and  
Ref.\ \cite{B112200} for the four-loop case $(q=4,x=1,r=0)$.}
For the diagram in Fig.\ (\ref{Example}), this case corresponds to $z=1$,
which is a singular point for the differential reduction algorithm.
The question is then how to find a smooth limit at this point.
Let us recall \cite{hyper} that the hypergeometric function 
(\ref{hypergeometric}) converges at $z=1$ if 
${\rm Re}\left(\Sigma b_j - \Sigma a_i \right)>0$.
In this way, if the hypergeometric function on the l.h.s. of 
Eq.\ (\ref{decomposition}) is well-defined at $z=1$, 
a smooth limit of the differential reduction exists. It is now a technical 
problem to rewrite the r.h.s. of Eq.\ (\ref{decomposition}) in terms of the 
variable $x=1-z$.  It is well-known, however, that for $p \geq 3$, the 
hypergeometric function $_pF_{p-1}$ is not expressible in terms of 
hypergeometric functions of the same type in the neighborhood of $z=1$ 
(see Ref.\ \cite{analytical}). 

One approach to this problem is to construct the all-order 
$\ep$-expansion\footnote{See Ref.\ \cite{skorokhodov} for 
another technique of evaluating hypergeometric functions at $z=1$.}
in terms of functions which are defined for the entire range $0 \leq z \leq 1$. 
The problem is then solved at each order in $\ep$. 
The hyperlogarithms \cite{LD} belong to that class. Unfortunately, at present, 
the necessary theorems on the all-order $\ep$-expansion are proven only for 
special sets of parameters\cite{nested,KWY,Oi}. 
Let us recall some results from Ref.\ \cite{KWY}.
There are bases (sets of parameters) for which the all-order $\ep$-expansion 
of a hypergeometric function has the form (see Ref.\ \cite{KWY})
\begin{equation}
_{p}F_{p-1}(\vec{A}+\ep\vec{a};\vec{B}+\ep\vec{b};z) = 
 C(\vec{A},\vec{a},\vec{B},\vec{b},z) \sum_{j=0}^\infty \ep^j 
\sum_{\vec{J},k,\vec{s}=1}
c_{\vec{s}}(\vec{a},\vec{b})
\Li{\vec{s}}{ \vec{\lambda_q}^{\vec{J}}, \lambda_q^k \xi} \;,
\label{expansion}
\end{equation}
where 
$\sum_i s_i = j$, 
$ 1 \leq j_a, k \leq q$, and $q$ is integer number, 
the coefficients $c_{\vec{s}}(\vec{a},\vec{b})$ are polynomials in the 
parameters $\{a_j\}$ and $\{b_k\}$, 
$\lambda_q$ is primitive $q^{\rm th}$-root of unity, 
$\Li{\vec{s}}{\vec{\lambda_q}^{\vec{J}}, \lambda_q^k \xi}$ are 
hyperlogarithms, $\vec{\lambda_q}^{\vec{J}}$ is short-hand for
$\lambda_q^{j_1-j_2}, \lambda_q^{j_2-j_3}, \cdots,  \lambda_q^{j_{k-1}-j_k}$,
$\xi$ is a variable related algebraically\footnote{For completeness, we 
note that $\xi$ can take the explicit
forms $\xi_{1,2,3} = z^\frac{1}{q}, (1-z)^\frac{1}{q},  
\left( \frac{z}{z-1}\right) ^\frac{1}{q}$
where $q$ is an integer. 
Under $z \to 1-z$, these variables transform as 
$\xi_{1,2} \to  \xi_{2,1}$ and $\xi_3 \to \frac{1}{\xi_3}$. }
to $z$, and  $C(\vec{A},\vec{a},\vec{B},\vec{b},z)$ is a 
polynomial.  Thus, at each order in the $\ep$-expansion, only 
hyperlogarithms of a single weight are generated.
The definition of the hyperlogarithm as an iterated integral over any 
rational function 
\begin{eqnarray}
I(z;a_k, a_{k-1},\ldots , a_1) & = & 
\int_0^{z} \frac{dt_k}{t_k-a_k}
\int_0^{t_{k}} \frac{dt_{k-1}}{t_{k-1}-a_{k-1}}
\cdots 
\int_0^{t_{2}} \frac{dt_1}{t_1-a_1} 
\nonumber \\ & = & 
\int_0^{z} \frac{dt}{t-a_k}
I(t;a_{k-1},\ldots , a_1)
\;,
\label{I}
\end{eqnarray}
where the $a_j$ are arbitrary numbers, together with the structure of 
the expansion (\ref{expansion}), show that the 
transformation $z \to 1-z$ results in functions of the same structure.  
In this way, we can construct the necessary limiting procedure without 
detailed knowledge about the relationship between the functions
$
{}_pF_{p-1}(\vec{a};\vec{b};z) 
$
and
$
{}_pF_{p-1}(\vec{a};\vec{b};1-z) .
$
For practical applications to diagrams presently of interest, a few 
coefficients of the $\ep$-expansion suffice, and these have been implemented 
in several existing packages \cite{packages}.

\begin{footnotesize}

\end{footnotesize}

\begin{thebibliography}{99}
\bibitem{url} Presentation: S.A. Yost, \\
\url{http://ilcagenda.linearcollider.org/contributionDisplay.py?contribId=76&sessionId=18&confId=2628}
\bibitem{acat08}
M.Yu.~Kalmykov, V.V.Bytev, B.A.~Kniehl, B.F.L.~Ward,  S.A.~Yost,
Proc. ACAT 2008 [arxiv:0901.4716].

\bibitem{numerator}
A.I.~Davydychev,
Phys.\ Lett.\  {\bf B263} (1991) 107; 
O.V.~Tarasov, 
Phys.\ Rev.\ D {\bf 54} (1996) 6479.

\bibitem{Smirnov}
V.A.~Smirnov,
\emph{Feynman Integral Calculus}, (2006, Springer, Berlin).

\bibitem{Takayama}
N.~Takayama,
Japan J.\ Appl.\ Math.\  {\bf 6}  (1989) 147.

\bibitem{Rainville}
E.D.~Rainville,
Bull.\ Amer.\ Math.\ Soc.\  {\bf 51} (1945) 714.

\bibitem{BKK}
V.~Bytev, M.Yu.~Kalmykov, B.~Kniehl, 
\emph{HYPERDIRE}, in press. 

\bibitem{quark08}
M.Yu.~Kalmykov,
JHEP {\bf 0604} (2006) 056; \\
M.Yu.~Kalmykov, B.A.~Kniehl, B.F.L.~Ward, S.A.~Yost,
Proc. of Quarks-08, [arXiv:0810.3238].

\bibitem{BD}
E.E.~Boos, A.I.~Davydychev,
Theor.\ Math.\ Phys.\  {\bf 89} (1991) 1052.

\bibitem{david}
D.J.~Broadhurst,
arXiv:hep-th/9604128.

\bibitem{hyper}
A.~Erdelyi (Ed.), {\it Higher Transcendental Functions}, vol.1 (McGraw-Hill, New York, 1953).

\bibitem{B112200}
Y.~Schr\"oder, A.~Vuorinen,
JHEP {\bf 0506} (2005) 051; \\
K.G.~Chetyrkin, J.H.~Kuhn, P.~Mastrolia, C.~Sturm,
Eur.\ Phys.\ J.\  {\bf C40} (2005) 361; \\
K.G.~Chetyrkin, M.~Faisst, C.~Sturm, M.~Tentyukov,
Nucl.\ Phys.\   {\bf B742} (2006) 208; \\
B.A.~Kniehl, A.V.~Kotikov,
Phys.\ Lett.\  {\bf B638} (2006) 531;
{\it ibid.} {\bf 642} (2006) 68; \\
B.A.~Kniehl, A.V.~Kotikov, A.I.~Onishchenko, O.L.~Veretin,
Phys.\ Rev.\ Lett.\  {\bf 97} (2006) 042001.

\bibitem{analytical}
N.E.~Norlund, 
Acta Math. {\bf 94} (1955) 289; 
P.O.M.~Olsson, 
J.\ Math.\ Phys.\  {\bf 7} (1966) 702; \\
W.~B\"uhring, 
SIAM J.\ Math.\ Anal.\  {\bf 19}  (1988) 1249.

\bibitem{skorokhodov}
S.L.~Skorokhodov,
Comput.\ Math.\ Math.\ Phys.\  {\bf 41}  (2001) 1718; 
{\it ibid.} {\bf 44}  (2004) 1102;
Program.\ Comput.\ Software  {\bf 29}  (2003) 75; 
{\it ibid.}  {\bf 30}  (2004) 150;
Comput.\ Math.\ Math.\ Phys.\  {\bf 45} (2005) 550; \\
A.I.~Bogolubsky, S.L.~Skorokhodov, 
Program.\ Comput.\ Software  {\bf 32} (2006) 145.

\bibitem{DK}
A.I.~Davydychev, M.Yu.~Kalmykov,
Nucl.\ Phys.\ Proc.\ Suppl.\  {\bf 89} (2000) 283; 
Nucl.\ Phys.\  {\bf B605} (2001) 266.

\bibitem{LD}
J.A.~Lappo-Danilevsky,
\emph{M\'emoires sur la th\'eorie des syst\'emes des \'equations diff\'erentielles lin\'eaires},
(Chelsea, New York, 1953); \\
A.B.~Goncharov, 
Math.\ Res.\ Lett.\  {\bf 4}  (1997) 617; 
Math.\ Res.\ Lett.\  {\bf 5}  (1998) 497; \\
E.~Remiddi, J.A.M.~Vermaseren,
Int.\ J.\ Mod.\ Phys.\  {\bf A15} (2000) 725; \\
 J.M.~Borwein, D.M.~Bradley, D.J.~Broadhurst, P.~Lisonek,
Trans.\ Am.\ Math.\ Soc.\  {\bf 353} (2001) 907; \\
T.~Gehrmann, E.~Remiddi,
Nucl.\ Phys.\  {\bf B601} (2001) 248; \\
J.~Vollinga, S.~Weinzierl,
Comput.\ Phys.\ Commun.\  {\bf 167} (2005) 177.

\bibitem{nested}
S.~Moch, P.~Uwer, S.~Weinzierl,
J.\ Math.\ Phys.\  {\bf 43} (2002) 3363; \\
S.~Weinzierl,
J.\ Math.\ Phys.\  {\bf 45} (2004) 2656.

\bibitem{KWY}
M.Yu.~Kalmykov, B.F.L.~Ward, S.~Yost,
JHEP {\bf 0702} (2007) 040; 
{\it ibid.} {\bf 0710} (2007) 048; 
{\it ibid.} {\bf 0711} (2007) 009; \\
S.A.~Yost, M.Yu.~Kalmykov, B.F.L.~Ward,
Proc. of ICHEP 2008 [arXiv:0808.2605]; \\
M.Yu.~Kalmykov, B.~Kniehl,
Nucl.\ Phys.\  {\bf B809}  (2009) 365.

\bibitem{Oi}
Shu Oi, 
[arXiv:0810.1829 (math.QA)].

\bibitem{packages}
J.~Fleischer, A.~V.~Kotikov, O.L.~Veretin,
Nucl.\ Phys.\   {\bf B547} (1999) 343; \\
S.~Weinzierl,
Comput.\ Phys.\ Commun.\  {\bf 145} (2002) 357; \\
M.Yu.~Kalmykov, O.~Veretin,
Phys.\ Lett.\  B {\bf 483} (2000) 315; \\
F.~Jegerlehner, M.Yu.~Kalmykov, O.~Veretin,
Nucl.\ Phys.\ {\bf B658} (2003) 49; \\
A.I.~Davydychev, M.Yu.~Kalmykov,
Nucl.\ Phys.\  {\bf B699} (2004) 3; \\
M.Yu.~Kalmykov,
Nucl.\ Phys.\ Proc.\ Suppl.\  {\bf 135} (2004) 280; \\
S.~Moch, P.~Uwer,
Comput.\ Phys.\ Commun.\  {\bf 174} (2006) 759; \\
T.~Huber, D.~Ma\^{\i}tre,
Comput.\ Phys.\ Commun.\  {\bf 175} (2006) 122; 
{\it ibid.} {\bf 178} (2008) 755.
\end{thebibliography}
\end{document}